\documentclass[aps,prl,reprint,groupedaddress]{revtex4-1}
\usepackage{graphicx} 
\usepackage{dcolumn} 
\usepackage{bm}
\usepackage{hyperref}

\begin{document}

\title{Strangeness production and hypernuclear formation in proton and antiproton induced reactions}
\author{Zhao-Qing Feng}
\email{fengzhq@scut.edu.cn}

\affiliation{School of Physics and Optoelectronics, South China University of Technology, Guangzhou 510640, China}

\date{\today}

\begin{abstract}
Strange particles and hyperfragments in collisions of antiprotons and protons on nuclei have been investigated systematically within a microscopic transport model. The hyperons are produced from the annihilation in antibaryon-baryon collisions and strangeness exchange process in antiproton induced reactions. A coalescence approach is used for constructing the primary hyperfragments in phase space and the statistical model is modified for describing the decay of hyperfragments via evaporating hyperon, neutron, charged particles etc, in which the shell effect, binding energy and root-mean-square radii are taken into account. It is found that the influence of the hyperon-nucleon interaction on the free $\Lambda$ and $\Xi^{-}$ production is negligible. However, the large hyperfragment yields are obvious with the attractive potential. The production of double strangeness hyperfragments are reduced below 1$\mu b$ in comparison to the yields of $\Lambda$-hyperfragments with the cross sections of 0.05-0.1 mb in the antiproton induced reactions on $^{63}$Cu at the incident momenta of 1-5 GeV/c. The light hyperfragments are formed in the dynamical fragmentation process. The energy dependence of hyperfragment formation is weak once the incident energy above the threshold energy for the hyperon production.

\begin{description}
\item[PACS number(s)]
21.80.+a, 25.40.-h, 25.43.+t
\end{description}
\end{abstract}

\maketitle

\section{I. Introduction}

In past several decades, strangeness nuclear physics has been extensively investigated both in experiments and in theories on the issues of hypernuclear physics, nuclear equation of state (EOS), hadronic matter properties, chiral symmetry restoration etc. Inclusion of the strangeness degree of freedom in a nucleus extends the research activities in nuclear physics, in particular the of hypernucleus and kaonic nucleus, hyperon-nucleon and hyperon-hyperon interactions, probing the in-medium properties of hadrons \cite{Gi95,Fa04,Ha06, Le18,Ep15,Sa16,Ya19}. Since the first observation of $\Lambda$-hypernuclide in nuclear multifragmentation reactions induced by cosmic rays in 1950s \cite{Da53}, a remarkable progress has been obtained in terrestrial laboratories for producing hypernuclides via different reaction mechanism, such as hadron (pion, K$^{\pm}$, proton, antiproton) induced reactions, bombarding the atomic nucleus with high-energy photons or electrons, and fragmentation reactions with high energy heavy-ion collisions. The spectroscopy, lifetime, giant monopole resonance, cluster structure and decay modes of  hypernuclei were investigated \cite{Na14,Ga16,Lv18,Hi18}. Recently, the antihypernuclide $^{3}_{\overline{\Lambda}}\overline{H}$ was found by the STAR collaboration in relativistic heavy-ion collisions and its binding energy was measured for the first time for testing the CPT theorem \cite{Star}. With the establishment of new facilities in the world, e.g., PANDA (Antiproton Annihilation at Darmstadt, Germany) \cite{Pand} and Super-FRS/NUSTAR \cite{Sa12} at FAIR (GSI, Germany), NICA (Dubna, Russia) \cite{Nica}, JPARC (Japan) \cite{Ta12}, HIAF (high-intensity heavy-ion accelerator facility, China) \cite{Ya13} etc, the hypernuclear physics is to be extended on the topics of isospin degree of freedom (neutron-rich/proton-rich hypernuclei), multiple strangeness nucleus, antihypernucleus, high-density hadronic matter with strangeness. The hypernucleus production with the antiprotons and high-energy protons is attracted much attention at PANDA and HIAF, respectively. The dynamics of hypernuclei is expected and helpful for the management of detector system in experiments.

The kinematics and properties of hypernuclei are related to the reaction mechanism. The exotic hypernuclei with extremely isospin asymmetry might be created via heavy-ion collisions. However, the antiproton and antikaon induced reactions are favorable for producing the multiple strangeness hypernuclei, in particular in the domain of heavy mass. The dynamics of hypernuclei in the antiproton-nucleus collisions is associated with the hyperon production in the annihilation and meson-baryon collisions, the capture of hyperon in the fragmentation process. The formation mechanism of hypernuclei in heavy-ion collisions and hadron induced reactions has been extensively investigated by several approaches, i.e., the statistical multifragmentation model (SMM) \cite{Bo07,Bl19}, statistical approach with a thermal source \cite{An11} and microscopic transport models such as Giessen Boltzmann-Uheling-Uhlenbeck (GiBUU) \cite{Gt14,Gt16} and quantum molecular dynamics (QMD) \cite{Bo15,Fe16,Fv19}. The dynamical description of hypernucleus formation is still expected for evaluating the production cross section, cluster effect in the preequilibrium process, correlation of $\Lambda-\Lambda$ interaction etc.

In this work, the production mechanism of strange particles and hypernuclei in the proton and antiproton induced reactions are to be investigated within the Lanzhou quantum molecular dynamics (LQMD) transport model. The article is organized as follows. In Sec. II I give a brief description of the model. The calculated results and discussion are presented in Sec. III. Summary and perspective on the hypernuclear production are outlined in Sec. IV.

\section{II. Brief description of the theoretical approach}

In the past several years, the LQMD transport model was developed for investigating the density dependence of symmetry energy, isospin splitting of nucleon effective mass and in-medium effects of hadrons in heavy-ion collisions, spallation reactions induced by hadrons, annihilation mechanism in antiproton-nucleus collisions. Recently, the model is modified for describing the formation mechanism of hypernuclei in heavy-ion collisions and in antikaon induced reactions \cite{Fe20}. In the LQMD transport model, the dynamics of resonances with the mass below 2 GeV ($\Delta$(1232), N*(1440), N*(1535) etc), hyperons ($\Lambda$, $\Sigma$, $\Xi$) and mesons ($\pi$, $\eta$, $K$, $\overline{K}$, $\rho$, $\omega$) is described by coupling the hadron-hadron collisions and rescattering processes via the decay reproduction of resonances, baryon-antibaryon annihilation reactions, meson-baryon and baryon-baryon collisions \cite{Fe11,Fe18}.
The temporal evolutions of nucleons and nucleonic resonances are described by Hamilton's equations of motion under the self-consistently generated two-body and three-body interaction potential with the Skyrme-like force. The one-body potentials of kaons (antikaons) and hyperons are used for transportation in nuclear medium and evaluated by the chiral effective Lagrangian approach and relativistic mean-field theories \cite{Fe13,Fe15}, respectively. The optical potential of hyperon is written as
\begin{equation}
V_{Y}(\textbf{p},\rho)=\omega_{Y}(\textbf{p},\rho)-\sqrt{\textbf{p}^{2}+m^{2}},
\end{equation}
in which the in-medium energy is derived from
\begin{equation}
\omega_{Y}(\textbf{p}_{i},\rho_{i})=\sqrt{(m_{H}+\Sigma_{S}^{H})^{2}+\textbf{p}_{i}^{2}} + \Sigma_{V}^{H}.
\end{equation}
The self-energies of hyperons are assumed to be two thirds of that experienced by nucleons, namely $\Sigma_{S}^{\Lambda}= 2 \Sigma_{S}^{N}/3$, $\Sigma_{V}^{\Lambda}= 2\Sigma_{V}^{N}/3$, $\Sigma_{S}^{\Xi}= \Sigma_{S}^{N}/3$ and $\Sigma_{V}^{\Xi}= \Sigma_{V}^{N}/3$. The nucleon scalar $\Sigma_{S}^{N}$ and vector $\Sigma_{V}^{N}$ self-energies are computed from the well-known relativistic mean-field model with the NL3 parameter ($g_{\sigma N}$=8.99, $g_{\omega N}$=12.45 and $g_{\rho N}$=4.47) \cite{La97}.  The values of optical potentials at saturation density are -32 MeV and -16 MeV for $\Lambda$ and $\Xi$, respectively. A weakly repulsive $\Sigma$N potential with 12 MeV at saturation density is used by fitting the calculations of chiral effective field theory \cite{Pe16}. The hyperon-nucleon interaction potentials will influence the dynamics of hyperons in nuclear medium, i.e., kinetic energy spectra, emission anisotropy etc. Furthermore, the bound states to form nuclear fragments and hypernuclei are modified by the potential.

The nuclear dynamics induced low-energy antiprotons has been extensively investigated within the LQMD model, i.e., the strange particle production, preequilibrium nucleon emission, fragmentation reactions etc. The antiproton-nucleon potential influences the reaction dynamics and the hyperfragment formation, which is calculated by performing the G-parity transformation of nucleon self-energies.
The optical potential of antiproton in nuclear medium is derived from the in-medium energy as
\begin{equation}
V_{\overline{p}}(\textbf{p},\rho)=\omega_{\overline{N}}(\textbf{p},\rho)-\sqrt{\textbf{p}^{2}+m^{2}}.
\end{equation}
The antinucleon energy in nuclear medium is evaluated by the dispersion relation as
\begin{equation}
\omega_{\overline{N}}(\textbf{p}_{i},\rho_{i})=\sqrt{(m_{N}+\Sigma_{S}^{\overline{N}})^{2}+\textbf{p}_{i}^{2}} + \Sigma_{V}^{\overline{N}}
\end{equation}
with $\Sigma_{S}^{\overline{N}}=\xi\Sigma_{S}^{N}$ and $\Sigma_{V}^{\overline{N}}=-\xi\Sigma_{V}^{N}$ with $\xi$=0.25. The strength of the optical potential $V_{\overline{N}}=-164$ MeV is obtained at the normal nuclear density $\rho_{0}$=0.16 fm$^{-3}$ by fitting the available experimental data of antiproton-nucleus scattering.

The hadron-hadron collisions in elastic, inelastic scattering and charge (strangeness) exchange reactions are performed with a Monte Carlo procedure and weighted by Pauli blocking of the final states. The choice of reaction channel is randomly performed by the weight of the channel probability, which is evaluated by comparing the channel to the total cross section. All possible channels of pseudoscalar mesons and hyperons in meson-baryon and baryon-baryon collisions are included for describing heavy-ion collisions and hadron (proton, $\pi^{\pm}$, K$^{\pm}$) induced reactions. To treat the antiproton-nucleus collisions, the annihilation channels, charge-exchange reactions (CEX), elastic (EL) and inelastic scattering with antibaryons and hyperon with multiple strangeness are included as follows \cite{Fe16},
\begin{eqnarray}
&&  \overline{B}B \rightarrow \texttt{annihilation}(\pi,\eta,\rho,\omega,K,\overline{K},\eta\prime,K^{\ast},\overline{K}^{\ast},\phi),
\nonumber  \\
&&  \overline{B}B \leftrightarrow \overline{B}B (\texttt{CEX, EL}),   \overline{N}N \leftrightarrow \overline{N}\Delta(\overline{\Delta}N), \overline{B}B \leftrightarrow \overline{Y}Y,        \nonumber  \\
&&  \overline{B}B \leftrightarrow \overline{\Xi}\Xi, \overline{K}B \leftrightarrow K\Xi, YY \leftrightarrow N\Xi, \overline{K}Y \leftrightarrow \pi\Xi
\end{eqnarray}
Here the B strands for a nucleon and $\Delta$(1232), Y($\Lambda$, $\Sigma$), $\Xi(\Xi^{0,-})$, K(K$^{0}$, K$^{+}$) and $\overline{K}$($\overline{K^{0}}$, K$^{-}$). The overline of B (Y) means its antiparticle. Mesons are the main products in the antiproton induced reactions. Besides the strangeness exchange reactions such as $\overline{K}N \rightarrow \pi Y$, hyperons are also contributed from the meson induced reactions $B\pi(\eta) \leftrightarrow YK$. The main products are pions with the wide energy spectra in the antiproton induced reactions. Therefore, the hypernuclei might be produced with the low-energy antiproton beams.

The fragments formed in the hadron induced reactions are mainly distributed in the target-like region. Once a hyperon is created, it might be captured by surrounding nuclear fragments to hypernuclei. The fragments are recognized with a coalescence model in phase space, in which the nucleons at freeze-out in nuclear collisions are considered to belong to one cluster with the relative momentum smaller than $P_{0}$ and with the relative distance smaller than $R_{0}$ (here $P_{0}$ = 200 MeV/c and $R_{0}$ = 3 fm). Actually, the influence of the coalescence parameters on the final fragments is small because the lager coalescence distance increases the excitation energy of primary fragment and enables the de-excitation process. The root-mean-square radii of the constructed fragment is checked with the normal radii-mass formula. The excitation energy is evaluated as the difference of binding energies between the excited fragment and the Bethe-Weiz\"{a}cker mass formula for nuclear fragments. The generalized mass formula with SU(6) symmetry breaking is used for calculating the hypernuclear binding energy \cite{Sa06}. The reaction system evolves until the excitation energy of 3\emph{A} MeV with the mass number of fragment \emph{A}, at which the nonequilibrium transportation is stopped and particle evaporation is dominant. The de-excitation process of the nuclear fragment and hyperfragments is described by the GEMINI code \cite{Ch88}, in which the channels of $\gamma$, light complex clusters (n, p, $\alpha$ etc) and binary fragments are selected by the Monte Carlo procedure via the decay width. The decay widths of light particles with Z$\leq$2 and the binary decay are calculated by the Hauser-Feshbach formalism \cite{Ha52} and transition state formalism \cite{Mo75}, respectively. The hyperon decay width is also evaluated by the Hauser-Feshbach approach the phenomenological hyperon binding energy. In the LQMD model, the binding energy of the primary fragment is calculated by the internal motion energy and interaction potential as
\begin{eqnarray}
E_{B}(Z_{i},N_{i})= &&\sum_{j}\sqrt{p_{j}^{2}+m_{j}^{2}}-m_{j}          \nonumber \\
&& +\frac{1}{2}\sum_{j,k,k\neq j}\int f_{j}(\textbf{r},\textbf{p},t)f_{k}(\textbf{r}^{\prime},\textbf{p}^{\prime},t)  \nonumber \\
&& \times v(\textbf{r},\textbf{r}^{\prime},\textbf{p},\textbf{p}^{\prime})    d\textbf{r}d\textbf{r}^{\prime} d\textbf{p}d\textbf{p}^{\prime}               \nonumber \\
&& +\frac{1}{6}\sum_{j,k,l}\sum_{k\neq j, k\neq l,j\neq l}
\int f_{j}(\textbf{r},\textbf{p},t) f_{k}(\textbf{r}^{\prime},\textbf{p}^{\prime},t)                            \nonumber \\
&& \times  f_{l}(\textbf{r}^{\prime\prime},\textbf{p}^{\prime\prime},t) v(\textbf{r},\textbf{r}^{\prime},\textbf{r}^{\prime\prime},
\textbf{p},\textbf{p}^{\prime},\textbf{p}^{\prime\prime})             \nonumber \\
&& \times d\textbf{r}d\textbf{r}^{\prime}d\textbf{r}^{\prime\prime} d\textbf{p}d\textbf{p}^{\prime}d\textbf{p}^{\prime\prime},
\end{eqnarray}
where the $\textbf{r}, \textbf{p}$ are the nucleon position in the center of mass of the $i-$th fragment $(Z_{i}, N_{i})$. The phase space density $f(\textbf{r},\textbf{p},t)$ is calculated with the QMD wave function. We count the binding energy of hyperfragment with $E_{B}(Z_{i},N_{i},N_{Y})=E_{B}(Z_{i},N_{i})+\sum_{j=1}^{N_{Y}}\omega(\textbf{p}_{j},\rho_{j})-m_{H}$ with the $Z_{i}$, $N_{i}$ and $N_{Y}$ being the proton, neutron and hyperon numbers, respectively. It is noticed that the combined approach of the coalescence method and statistical model is available for the medium and heavy fragments because of the reliable estimation of binding energy. In the hadron induced reactions, the fragments are distributed in the target-like region and the statistical decay is implemented only for the fragments with the mass number $A>12$. The emissions of light clusters are complicated in the nuclear collisions, which are related with the dynamical recognition via the nucleon-nucleon scattering and the structure effect at freeze out. Recently, a new approach FRIGA (Fragment Recognition in General Application) was proposed for identifying the nuclear fragments and hypernuclei in heavy-ion collisions \cite{Fv19}, in which the shell effect and odd-even effect are taken into account in recognizing the fragments.

\section{III. Results and discussion}

Mesons, resonances and hyperons produced in heavy-ion collisions and hadron-induced reactions are modified in the nuclear medium. The elementary cross section, decay width, effective mass, energy spectra, invariant decay channels etc have been extensively investigated \cite{Fr07,Ha12}. The dynamics of hyperons is significant in the hyperfragment formation, which has been investigated for extracting the high-density equation of state and the properties of hadronic matter formed in heavy-ion collisions. Shown in Fig. 1 is the influence of the optical potentials on the production of hyperons $\Lambda$ and $\Xi^{-}$ in collisions of $\overline{p}$ on $^{63}$Cu at the beam momentum of 3 GeV/c. The $\Xi^{-}$ production is strongly suppressed by the 3-order magnitude in comparison to the $\Lambda$ yields. The $\Lambda$-nucleon potential slightly change the spectra. The $\Xi^{-}$ yields are enhanced with the potential because of the reduction of threshold energy. Although the $\Lambda$ potential increases its production, the reabsorption process $\pi \Lambda\rightarrow \overline{K}N$ weakens the $\Lambda$ production owing to the abundant pions from the antiproton annihilation.

\begin{figure*}
\begin{center}
{\includegraphics*[width=0.8\textwidth]{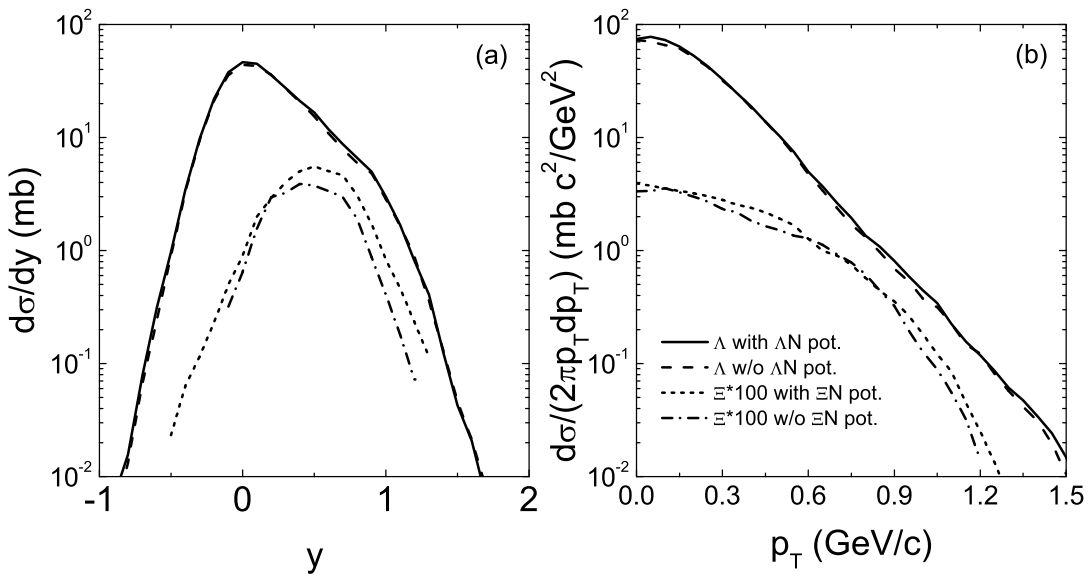}}
\end{center}
\caption{Rapidity and transverse momentum spectra of $\Lambda$ and $\Xi^{-}$ produced in the antiproton induced reactions on $^{63}$Cu at the beam momentum of 3 GeV/c. }
\end{figure*}

The strange particles produced in the antiproton induced reactions are associated with the annihilation and secondary collisions.  At the incident momentum above its threshold energy, e.g., the reactions $\overline{N}N\rightarrow \overline{\Lambda}\Lambda$ (p$_{\texttt{threshold}}$=1.44 GeV/c) and $\overline{N}N\rightarrow \overline{\Xi}\Xi$ (p$_{\texttt{threshold}}$=2.62 GeV/c), the production of hyperons are attributed from the direct reaction (annihilation and creation of quark pairs, $u\overline{u} (d\overline{d})\rightarrow s\overline{s}$) and also from the secondary collisions, i.e., meson induced reactions $\pi (\eta, \rho, \omega) N\rightarrow KY$ and strangeness exchange reaction $\overline{K}N\rightarrow \pi Y (K\Xi)$. At the momentum below the threshold energy, the strangeness exchange reactions and pion-nucleon collisions dominate the hyperon production. Shown in Fig. 2 is the invariant mass spectra of hyperons $\Lambda$ and $\Xi^{-}$ in collisions of $\overline{p}$ on $^{63}$Cu at different momenta and on the targets of $^{40}$Ca, $^{63}$Cu, $^{124}$Sn and $^{197}$Au at the momentum of 3 GeV/c. The momentum dependence of hyperon yields is pronounced in the high kinetic energy region. The hyperons are contributed from the direct annihilation process at the incident momentum above 3 GeV/c. The strangeness exchange reactions are obvious with increasing the mass number of target nuclide owing to the more collision probability between the meson and nucleon. The advantage of antiproton induced reactions is the multichannel contributions for hyperon production, which is also available for hyperfragment formation.

\begin{figure*}
\begin{center}
{\includegraphics*[width=0.8\textwidth]{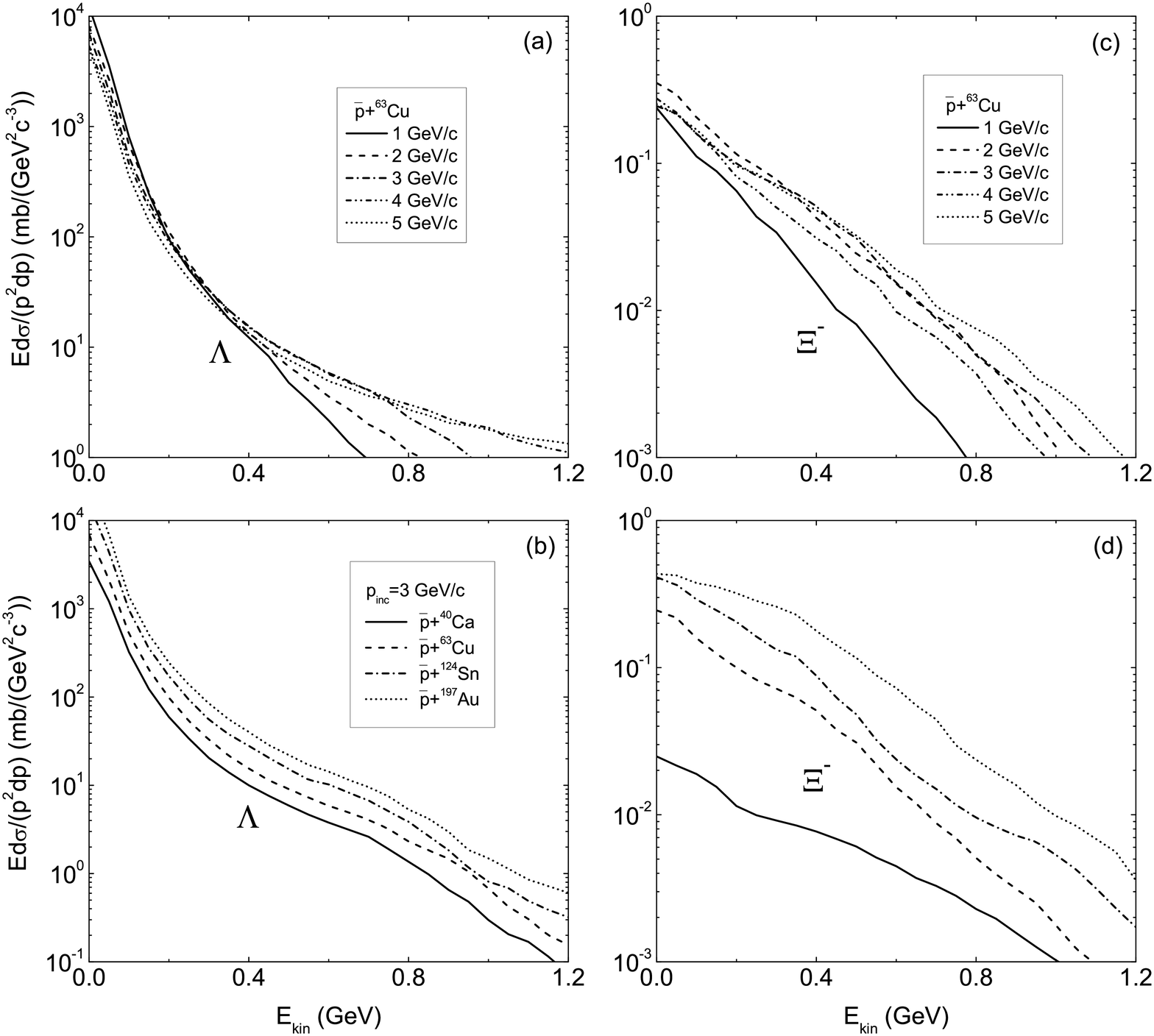}}
\end{center}
\caption{Inclusive spectra of $\Lambda$ and $\Xi^{-}$ produced in $\overline{p}+^{63}$Cu at different energies and on different targets at the beam momentum of 3 GeV/c.}
\end{figure*}

The target nucleus is heated by incoming antiproton via the annihilation reactions and the localized energy is released through the collisions of annihilation products and nucleons, which enables the formation of a highly excited nucleus. The energy deposition and spallation  mechanism in the antiproton induced reactions are to be investigated. Shown in Fig. 3 is the excitation energy distribution in collisions of $\overline{p}$ on $^{63}$Cu at the beam momenta of 1, 2, 3, 4 and 5 GeV/c in the left panel, and in the antiproton induced reactions on $^{40}$Ca, $^{63}$Cu, $^{124}$Sn and $^{197}$Au at 3 GeV/c. The collisions processes in the antiproton induced reactions undergo the elastic scattering, annihilation and secondary collisions with surrounding nucleons. The annihilation cross section decreases with increasing the antiproton energy. Therefore, the large excitation energy is released from the multiple collisions between pions and nucleons at the incident momentum of 1 GeV/c. The energy dissipation is obvious in the heavy target nuclide and leads to the high excitation energy. The fragments formed in the antiproton and proton induced reactions are positioned in the target-like region. The nuclear dynamics induced by antiprotons and protons is described by the LQMD model. The primary fragments are constructed with a coalescence model at freeze out stage in the nuclear collisions, in which the particle yields reach the equilibrium values. The primary fragments are highly excited and the de-excitation of the fragments are described by the statistical code. Shown in Fig. 4 is a comparison of the hyperon-nucleon potential and statistical decay on the hyperfragment production in the reaction p+$^{63}$Cu at 5 GeV/c. It is obvious that the attractive hyperon-nucleon potential is favorable the hyperfragment formation. The relative momentum between the hyperon and nuclear cluster is reduced with the potential, which enables the hyperon capture to a hypernuclide. Inclusion of the statistical decay in the primary fragments leads to the yield reduction of one-order magnitude.

\begin{figure*}
\begin{center}
{\includegraphics*[width=0.8\textwidth]{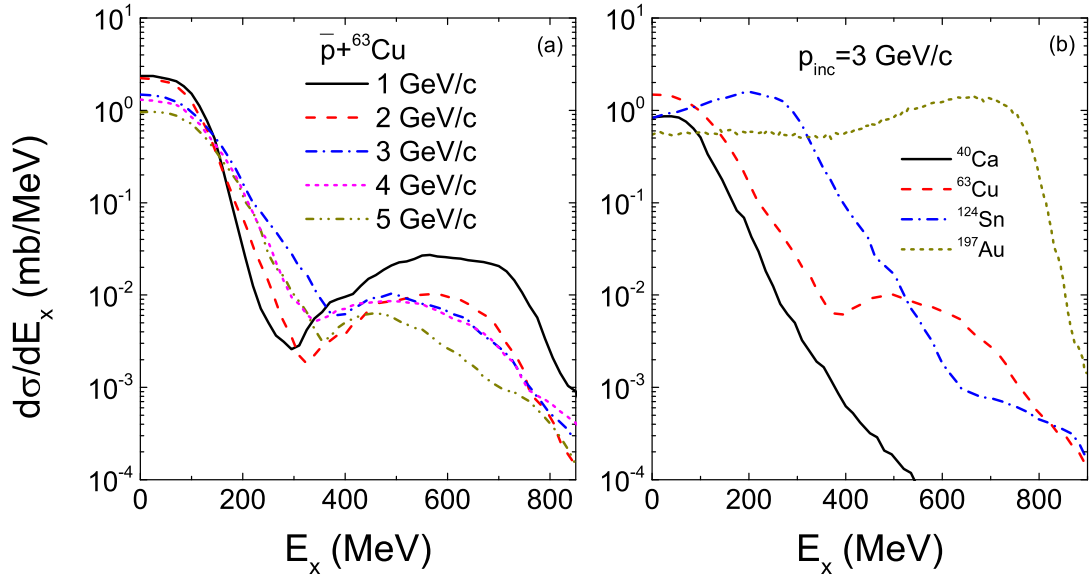}}
\end{center}
\caption{Excitation energy distribution in the antiproton induced reactions with different systems and incident momenta.}
\end{figure*}

\begin{figure*}
\begin{center}
{\includegraphics*[width=0.8\textwidth]{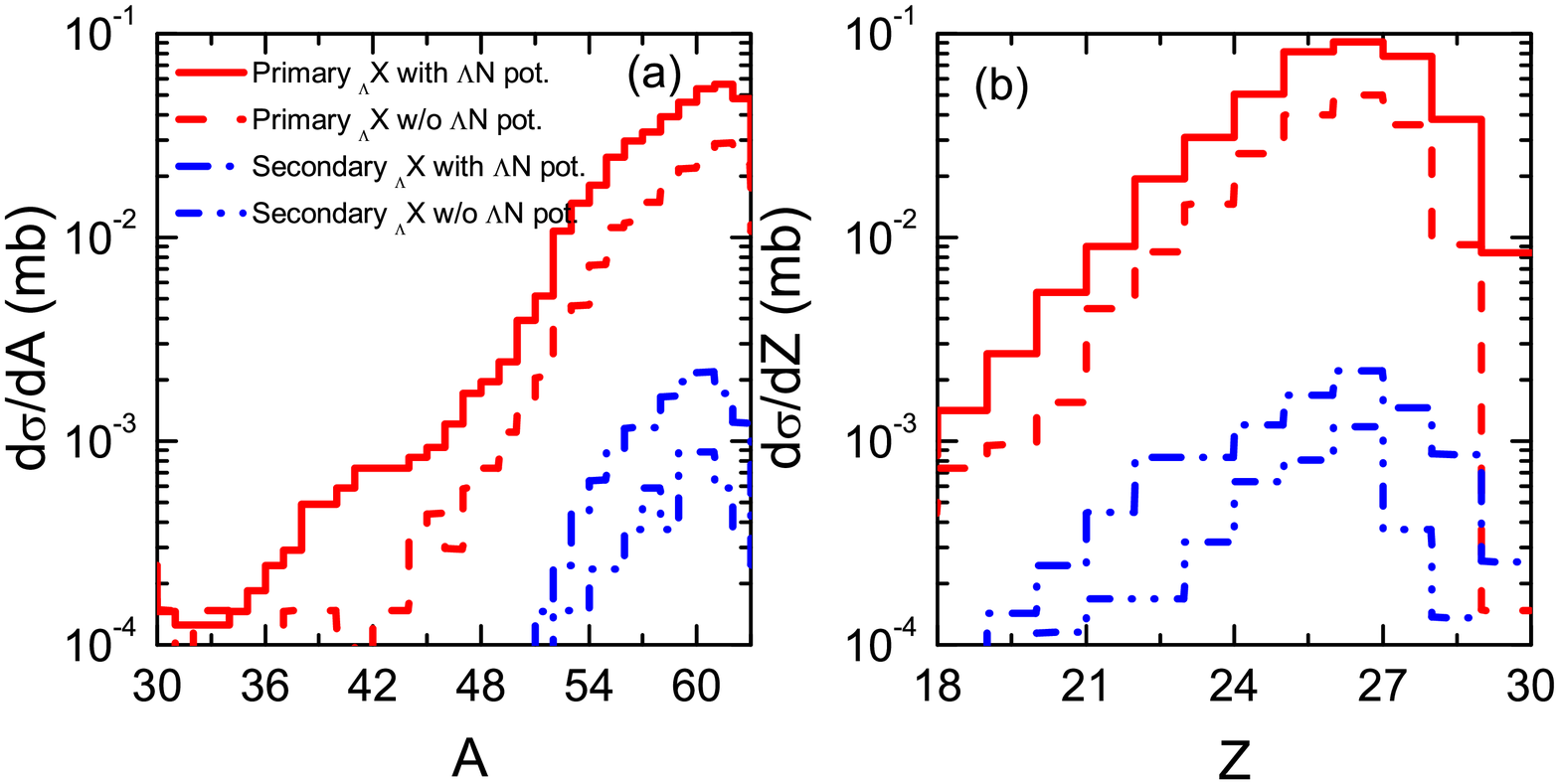}}
\end{center}
\caption{Influence of the statistical decay and hyperon-nucleon potential on the hyperfragment formation in the reaction of proton on $^{63}$Cu at an incident momentum of 5 GeV/c.}
\end{figure*}

The nuclear fragmentation reactions induced by antiprotons were extensively investigated in experiments with the LEAR (low-energy antiproton ring) facility at CERN, in which some topic issues were taken into account, i.e., the prequilibrium emissions of nucleons and clusters, the delayed fission from the decay of hypernuclei, particle production, in-medium effects of hadrons etc \cite{Ja93,Ho94,Ki96,Ja99}. Experimentally, the hypernuclei are reconstructed from the invariant mass spectra of $\pi^{-}$ and its decay fragments. The kinematics of hyperfragments is helpful for detector management in experiments, i.e., the rapidity distribution, transverse (kinetic energy) spectra, invariant energy, angular emission etc. The rapidity distributions of $\Lambda$-hyperfragments produced in the proton (5 GeV/c) and antiproton (2-5 GeV/c) induced reactions on $^{63}$Cu are calculated as shown in Fig. 5. It is obvious that the light fragments are formed in the spallation process and weakly depends on the incident momentum in the antiproton induced reactions. The spectra manifest a symmetric structure with antiprotons. The hypernuclear production is strongly suppressed with the proton-nucleus collisions.

\begin{figure*}
\begin{center}
{\includegraphics*[width=0.8\textwidth]{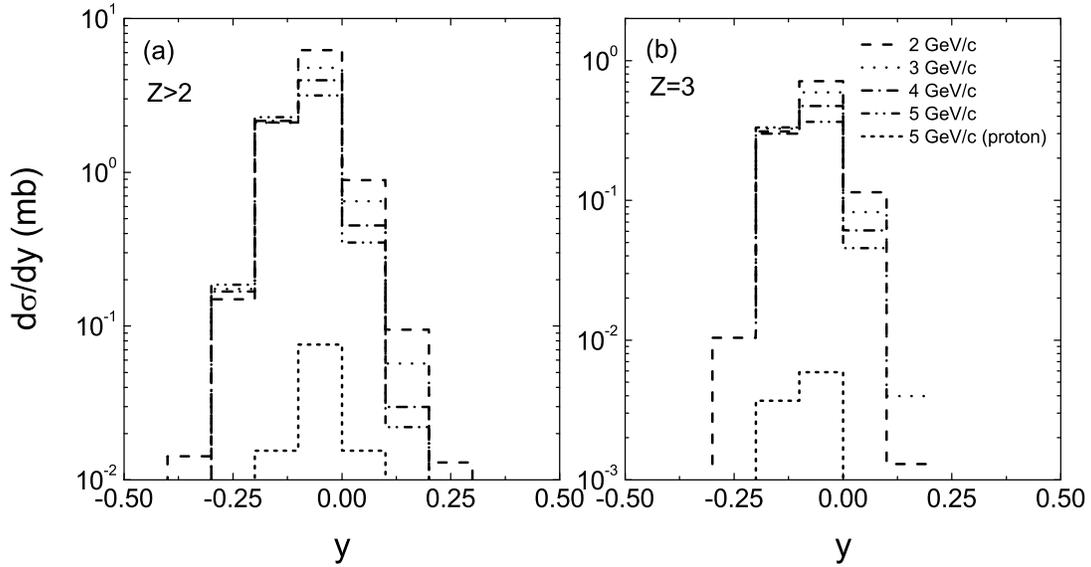}}
\end{center}
\caption{Rapidity distribution of hyperfragments with Z$>$2 and hyperlithium in collisions of antiprotons and protons on $^{63}$Cu.}
\end{figure*}

Once a hyperon is created inside the nucleus, the relative momentum between the hyperon and nucleon might be reduced to form a hypernucleus owing to the multiple collisions. It has advantages for creating heavy hypernuclei with the hadron induced reactions in comparison to heavy-ion collisions. The fragments and hyperfragments tend to be formed around the $\beta-$stability line in the fragmentation process. The hyperon is produced in the antiproton induced reactions at very low energies because of the contributions of meson-nucleon collisions and strangeness exchange process, e.g., $\pi N\rightarrow K\Lambda (p_{\texttt{threshold}}$=0.89 GeV/c), $\overline{K}N\rightarrow \pi \Lambda$. The direct annihilation dominates the hyperon production at high energy, $\overline{N}N\rightarrow \overline{\Lambda }\Lambda (p_{\texttt{threshold}}$= 1.44 GeV/c). Shown in Fig. 6 is a comparison of hyperfragments in the reaction $\overline{p}$ + $^{63}$Cu at different momenta. At the low incident momentum (1 GeV/c), more temporal dissipation enables the high excitation energy, which leads to the more nucleon or cluster emissions. Overall, the energy dependence of hyperfragment production is not obvious in the antiproton induced reactions. The properties of double strangeness hypernuclei are of importance in exploring the $\Lambda-\Lambda$ interaction, 3-body force of $\Lambda\Lambda$N state, short-range correlation etc. The double $\Lambda$ hypernucleus production at different incident momenta is calculated as shown in Fig. 7. The double strangeness hypernuclei are mainly contributed from the strangeness exchange reaction $N\Xi \rightarrow YY$. The dynamics of $\Xi$ is of significance for the hypernuclide formation. Once the hyperon $\Xi$ is created via the annihilation reaction $\overline{p}N \rightarrow \overline{\Xi}\Xi$ above the antiproton threshold momentum of $p_{\texttt{threshold}}$=2.62 GeV/c and the secondary collision $\overline{K}N \rightarrow K\Xi$ above the antikaon threshold momentum of $p_{\texttt{threshold}}$=1.04 GeV/c. A small distance between the hyperon and nucleon in phase space is favorable for bound hyperfragment. The maximal yields around 0.5-1 $\mu$b are found at the momenta of 2-5 GeV/c, which are feasible for experimental measurements at PANDA (Antiproton Annihilation at Darmstadt, Germany) in the near future.

\begin{figure*}
\begin{center}
{\includegraphics*[width=0.8\textwidth]{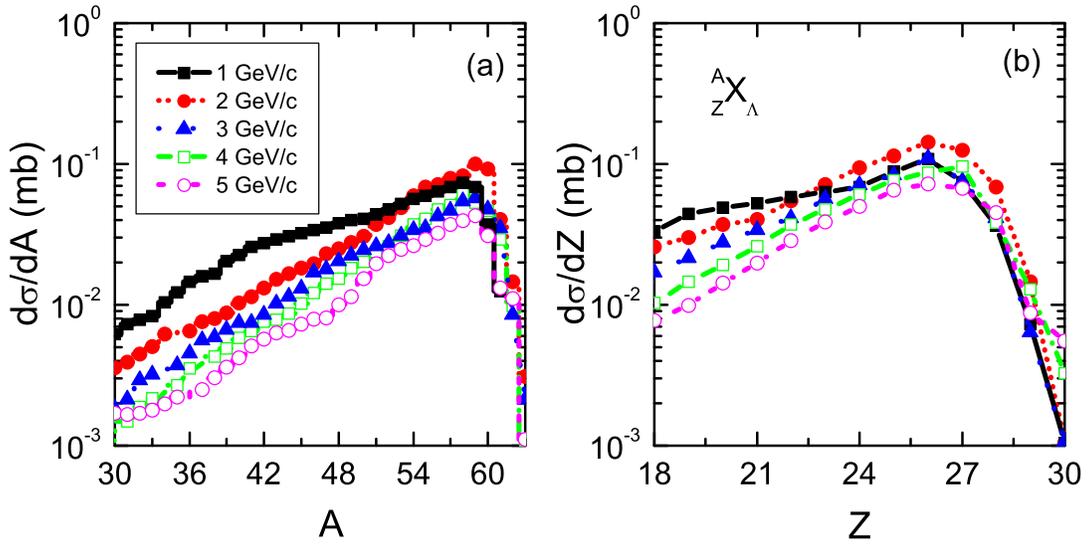}}
\end{center}
\caption{Incident energy dependence of the $\Lambda-$hyperfragments as functions of mass and charged numbers in collisions of $\overline{p}$+$^{63}$Cu, respectively.}
\end{figure*}

\begin{figure*}
\begin{center}
{\includegraphics*[width=16 cm]{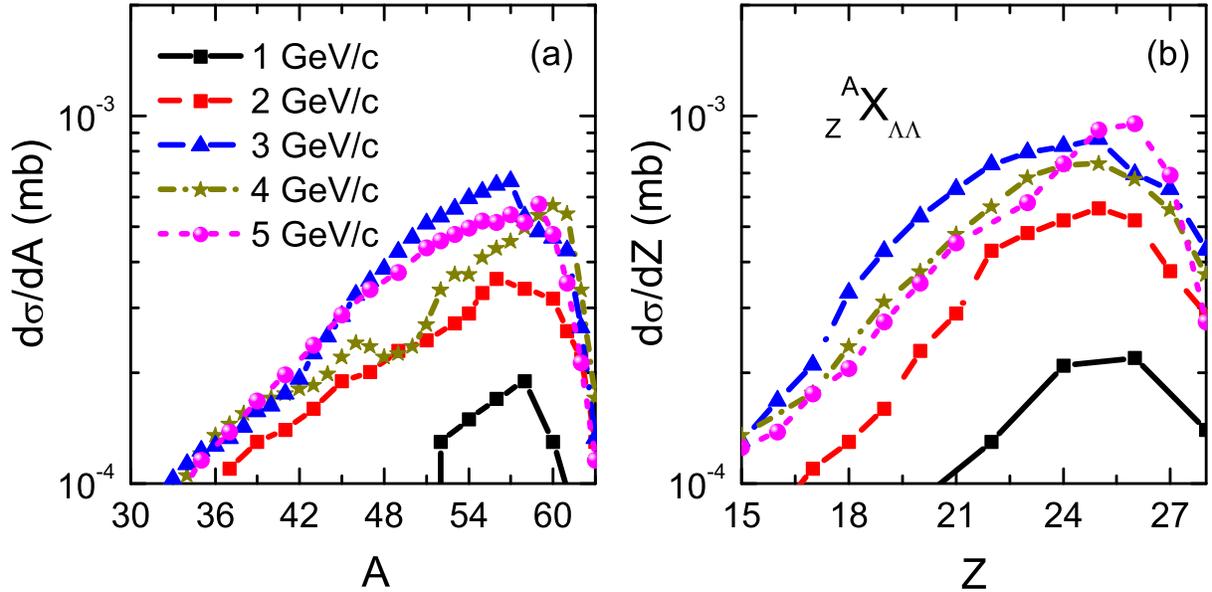}}
\end{center}
\caption{Hyperfragment production with double strangeness in the antiproton induced reactions at different energies.}
\end{figure*}

The spallation reactions induced by proton and antiproton have been extensively investigated with different models, which are associated with the inequilibrium process leading to the fragmentation and excitation of colliding system and with the decay modes via evaporating the $\gamma$-rays, particles and fission of heavy fragments. The SMM was applied for describing the multifragmentation and hypernucleus production. One needs to input the the mass, charge, hyperon numbers, and temperature for a composite system. Recently, the combined approaches of the SMM and transport models have been used for the hypernuclear production in heavy-ion collisions, i.e., UrQMD \cite{Bo17}, Dubna cascade model \cite{Su18}, GiBUU model \cite{Ga09}. A comparison of the LQMD and GiBUU for the antiproton induced reactions is shown in Fig. 8. The thermal source and hadron dynamics in collisions of $\overline{p}$ + $^{64}$Cu at the momentum of 5 GeV/c are provided by the GiBUU and the hypernuclide production in the fragmentation is described by the SMM \cite{Ga12}. The experimental data from the LEAR facility at CERN \cite{Ja93} are shown for a comparison. The mass distribution of nuclear fragments is nicely consistent with the data. It is noticed that the shapes of charge and mass distributions in the antiproton induced reactions weakly depend on the incident momentum, but strongly on the annihilation of antiproton in a nucleus \cite{Fe16}. The normal nuclear fragments are almost consistent in both calculations. However, the $\Lambda$- and $\Lambda\Lambda$- hyperfragments are reduced about the one-order magnitude with the LQMD in comparison with the estimation by GiBUU model. The possible reason is a part of hyperons escaping the nuclear composite system in the collisions. The large yields appear in the target-like regime and light fragments with the charge number Z$\leq$5 in the LQMD calculations. However, a platform is caused from the multifragmentatoin process in the combined GiBUU and SMM. More measurements of hyperfragments in the hadron (lepton) induced reactions and in heavy-ion collisions are expected for improving the theoretical description. The recognizing method is also significant for evaluating the nuclear fragments and hyperfragments, such as the fragment recognition in general application \cite{Fv19}. Besides the dynamical fragmentation and constructing the fragments at freeze out, the clusters in the multiple collisions also need to be taken into account in the transport models. Further investigation on the hyperfragment formation in the multiple collisions and dynamical recognizing method is in progress.

\begin{figure*}
\begin{center}
{\includegraphics*[width=16 cm]{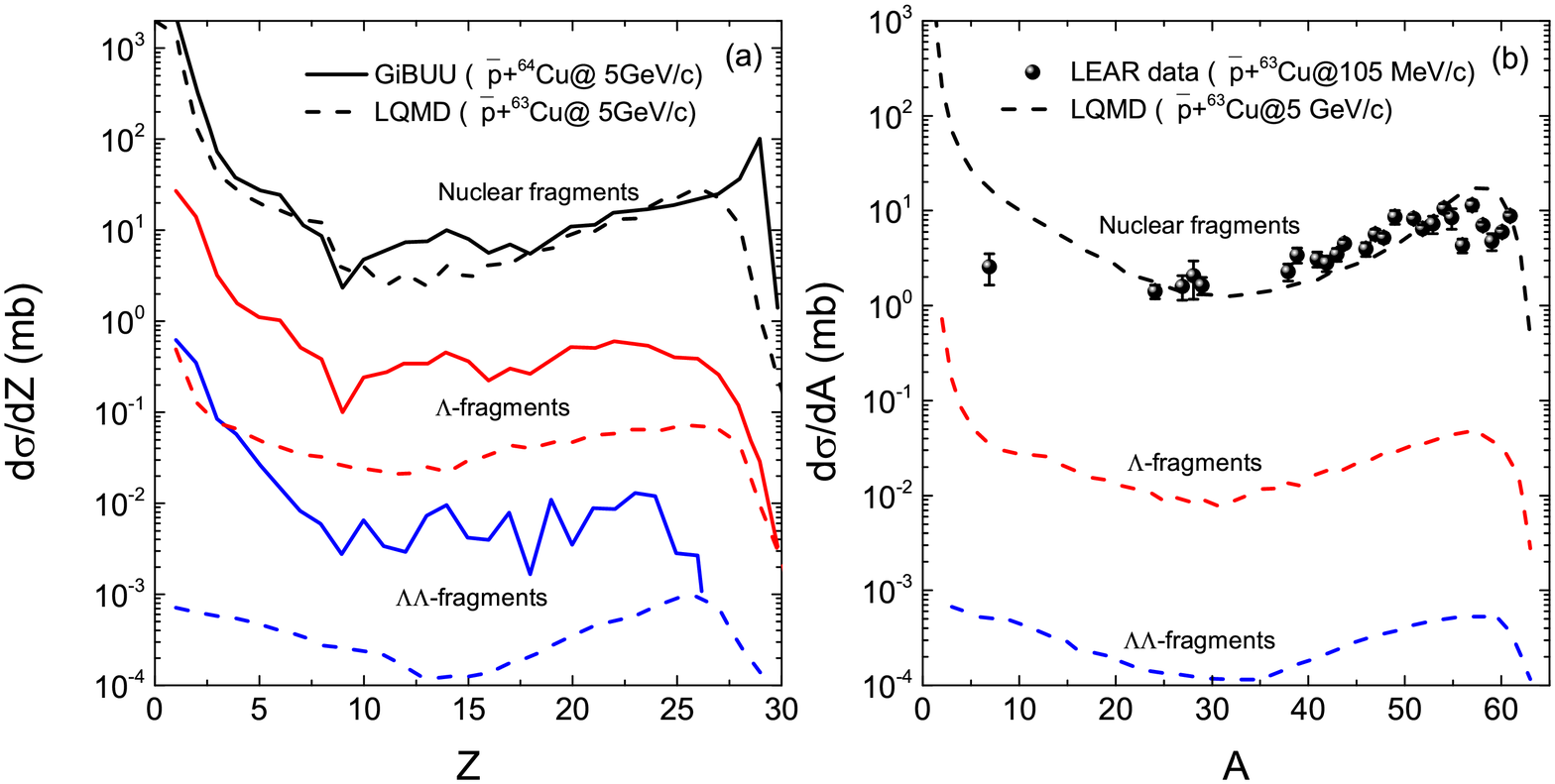}}
\end{center}
\caption{Mass and charge distributions of fragments produced in the $\overline{p}$ + $^{63}$Cu reaction at the incident momentum of 5 GeV/c. The calculations from GiBUU model $\overline{p}$ + $^{64}$Cu at 5 GeV/c \cite{Ga12} and the available data from LEAR facility at CERN \cite{Ja93} are shown for a comparison.}
\end{figure*}

\begin{figure*}
\begin{center}
{\includegraphics*[width=16 cm]{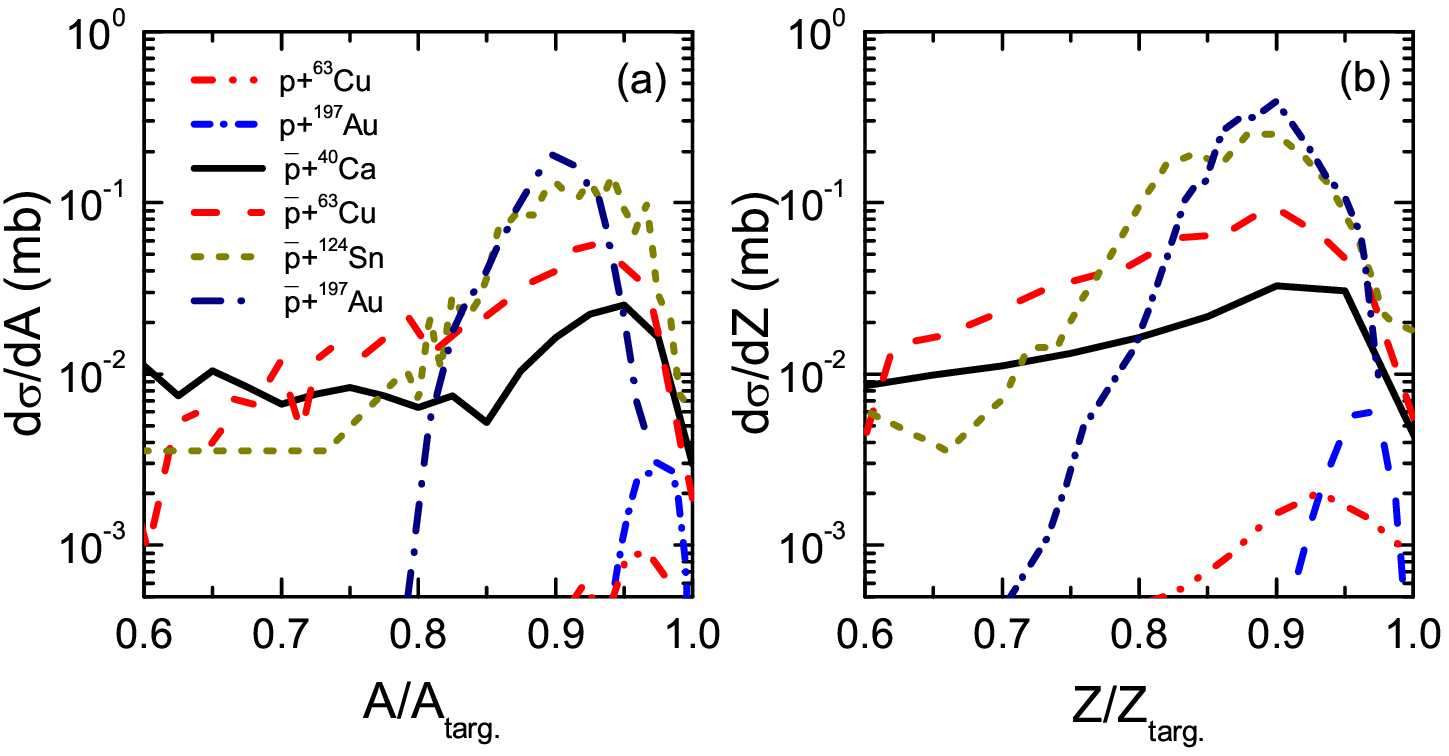}}
\end{center}
\caption{Comparison of $\Lambda-$hyperfragment production with proton and antiproton induced reactions on different targets.}
\end{figure*}

The formation of hyperfragments in the hadron induced reactions is related to the reaction system. The energy relaxation is pronounced in heavy target because of the multiple step collisions with surrounding nucleons, which influence the excitation magnitude and target fragmentation. Usually, the excitation energy increases with the mass of target nucleus and the fragmentation is also explosive. Shown in Fig. 9 is a comparison of the $\Lambda-$hyperfragments in the proton induced reactions on the targets $^{63}$Cu and $^{197}$Au at the momentum of 5 GeV/c, and antiprotons on $^{40}$Ca, $^{63}$Cu, $^{124}$Sn and $^{197}$Au at 3 GeV/c. The antiproton beams are favorable for creating the hypernuclei with the maximal cross section above 0.01 mb and increase the yields over two-order magnitude in comparison to the proton induced reactions. The hypernucleus production with the proton beams is feasible with the cross section above 1 $\mu$b. Both reactions induced by proton and antiproton beams manifest the spallation mechanism, in which the yields of intermediate fragments (IMF) are low. The fragments are mainly distributed in the target-like and light-mass domain. The experiments for hypernuclear physics with the high-energy protons are planning at the HIAF facility in the near future.

\section{IV. Conclusions}

In summary, the dynamics of strange particles and hypernuclear production in the proton and antiproton induced reactions have been investigated within the LQMD transport model. The hyperons $\Lambda$ and $\Xi^{-}$ are mainly created from the strangeness exchange reactions, meson-nucleon collisions and direct annihilations in collisions of antiproton on the target nucleus. The nucleon-nucleon direct collisions contribute the hyperon production in the proton induced reactions. The hyperfragments are formed in the target-like domain with the hadron induced reactions. The influence of the hyperon-nucleon potential on the hyperon energy spectra is negligible, but favorable for the hyperfragment formation. Inclusion of the statistical decay leads to the one-order magnitude reduction of hyperfragments. The production cross sections of double strangeness hypernuclei with the antiproton beams are found at the level of 1 $\mu$b, which are feasible for experiments at PANDA. The yields of hyperfragments are independent on the incident energy once above the threshold energies in the annihilation reactions for direct hyperon production. The $\Lambda$-hyperfragments with the high-energy proton beams might be measured at HIAF.

\section{Acknowledgements}

This work was supported by the National Natural Science Foundation of China (Projects No. 11722546 and No. 11675226) and the Talent Program of South China University of Technology.

\end{document}